\documentstyle[11pt,revpasp,psfig]{article}
 
\begin{document}
 
\title{ Molecular gas and the dynamics of galaxies }
\author{F. Combes}
\affil{Observatoire de Paris, DEMIRM,
61 Av. de l'Observatoire, F-75014, Paris,  France}
 
\begin{abstract}
In this review, I discuss some highlights of recent research
on molecular gas in galaxies; large-scale CO maps of nearby galaxies
are being made, which extend our knowledge on global properties, radial
gradients, and spiral structure of the molecular ISM. 
Very high resolution are provided
by the interferometers, that reveal high velocity gradients
in galaxy nuclei, and formation of embedded structures, like bars within bars.
 Observation of the CO and other lines in starburst galaxies have
questioned the H$_2$-to-CO conversion factor. 
Surveys of dwarfs have shown how the conversion factor depends on
metallicity. The molecular content is not deficient in galaxy clusters,
as is the atomic gas. Galaxy interactions are very effective to enhance
gas concentrations and trigger starbursts. Nuclear disks or rings
are frequently observed, that concentrate the star formation activity.
Since the density of starbursting galaxies is strongly increasing with
redshift, the CO lines are a privileged tool to follow evolution of 
galaxies and observe the ISM dynamics at high redshift: due to the
high excitation of the molecular gas, the stronger high-$J$ CO lines
are redshifted into the observable band, which facilitates the detection.

In the last years, progress has been very rapid in the domain of molecules
at high redshift, and we now know in better detail the molecular content
in several systems beyond $z=1$ and up to $z \sim 5$, either through
millimeter and sub-millimeter emission lines, continuum dust emission,
or millimeter absorption lines in front of quasars.
More and more systems are gravitationally lensed, which helps the
detection, but also complicates the interpretation.
The detection of all these systems
could give an answer about the debated question of the star-formation rate 
as a function of redshift. The maximum of star-formation rate, found
around $z=2$ from optical studies, could shift to higher $z$ if the most 
remote objects are hidden by dust.
\end{abstract}
 
 
\section{CO and H$_2$ content of galaxies}

Although the molecular component is a key parameter for the star formation
history and the evolution of galaxy disks, their H$_2$ content is
still very uncertain, mainly because the bulk of the gas is not seen
directly, but through questionable tracers, such as the CO lines.
 The wider surveys now available, together with the observations
of starbursts  and the more sensitive observations 
of low metallicity galaxies now have revealed how variable can be
the CO to H$_2$ conversion ratio, that was previously thought
constant within $\pm$ 50\% (Young \& Scoville 1991).
Also, the first surveys were oriented towards star forming galaxies,
selected from far-infrared (IRAS) samples, and those happened to
be rich CO emitters (according to the well established FIR-CO
correlation). When more galaxies are included, the derived 
H$_2$/HI mass ratio in galaxies becomes lower than 1, the previously
established value, to be around 0.2 in average (Casoli et al 1998). 

The dependence of the molecular content with type also has been refined;
the apparent H$_2$/HI mass ratio decreases monotonously from
SO to Sm galaxies, but the extremes were separated by a factor
20-30 (Young \& Knezek 1989), which is now reduced to about 10
(cf fig.1, Casoli et al 1998). This gradient towards the late-types might be
only due to a reduced CO emission, because of metallicity effects,
and not to an intrinsic reduction of the H$_2$ content. When only the more massive objects
are taken into account, this tendency with type does not appear:
there is no gradient of molecular fraction.
This supports the hypothesis that the gradient is due
to metallicity, which is correlated with total mass. 

\begin{figure}
\psfig{figure=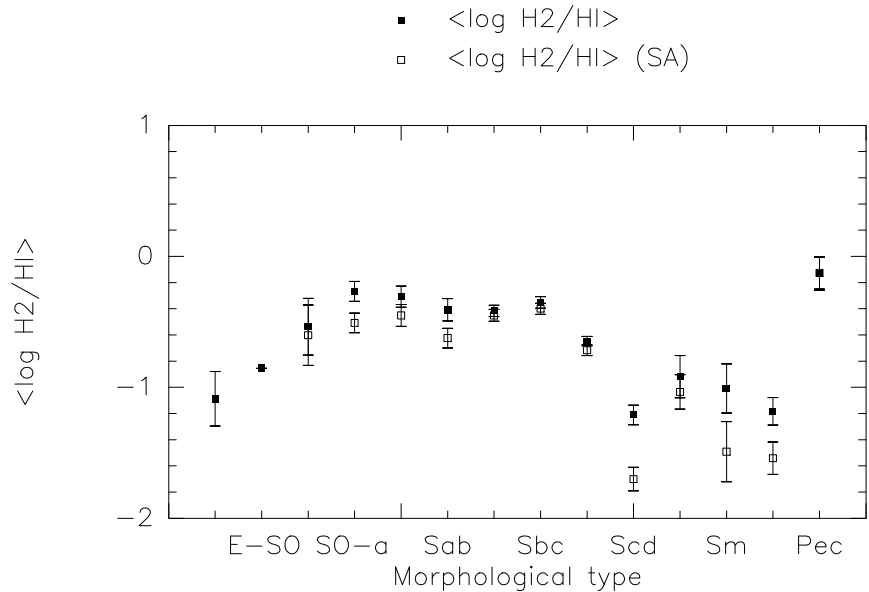,bbllx=6cm,bblly=1cm,bburx=18cm,bbury=18cm,width=12cm,angle=-90}
\caption{H$_2$/HI mass ratio in galaxies as a function of morphological
type; full squares: mean values with upper limits treated as detections;
empty squares:  mean values with upper limits taken into account using
the survival analysis (SA), from Casoli et al (1998).}
\end{figure}

The dependence of the CO-to-H$_2$ conversion factor X with metallicity
has been confirmed now in many galaxies. In the Small Magellanic Clouds,
 X could be 10 times higher than the standard value of 2.3 10$^{20}$ mol
cm$^{-2}$ (K.km/s)$^{-1}$ (Rubio et al 1993). The effect has been seen 
also in local group galaxies, such as M31, M33, NGC 6822 and IC10 (Wilson
1995). The physical explanation is complex, since the CO lines
are optically thick, but the size of the clouds (and therefore the
filling factor) decreases with metallicity, both due to direct 
CO abundance, and UV-phtotodissociation increased by the depletion of dust.
When the dust is depleted by a factor 20, there is only 10\% less H$_2$
but 95\% less CO (Maloney \& Black 1988).

Another tracer of the molecular gas has been widely used in recent years:
cold dust emission, through galaxy mapping with bolometers at 1.3mm.
 The technique is best suited to edge-on galaxies, since a nearby empty 
reference position must be frequently observed to eliminate atmospheric
gradients. One of the first observed, NGC 891, has a radial dust emission
profile exactly superposable to the CO-emission profile (Guelin et al 1993).
  At 3mm, the dust is completely optically thin, and therefore the 
emission is proportional to the dust abundance, or the metallicity Z.
The identity between the dust and CO emission profiles tend to confirm
the strong dependence of CO emission with Z.
In some galaxies, the CO emission falls off radially even faster than
the dust emission; it is the case of NGC 4565 or NGC 5907 
for example (cf fig 2, Neininger et al 1996).

\begin{figure}
\psfig{figure=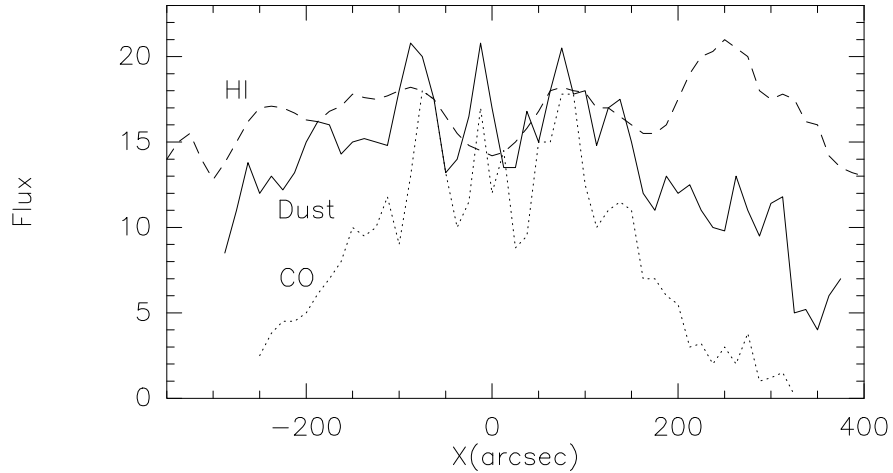,bbllx=6cm,bblly=1cm,bburx=18cm,bbury=18cm,width=12cm,angle=-90}
\caption{ Dust emission radial profile in NGC 4565, compared to the CO
and HI profiles, from Neininger et al (1996). 
}
\end{figure}

In those galaxies, where the CO emission falls faster than dust 
emission, the latter is correlated to the HI distribution (see also
the case of NGC 1068, Papadopoulos \& Seaquist 1999). This could be 
interpreted by two effects: there is an extended 
 diffuse molecular component, not visible in the CO lines, because 
the H$_2$ density is not sufficient to excite the CO lines,
or the CO abundance is depending non-linearly with metallicity,
i.e. decreases more than the dust, which abundance is linear
in Z. This might be the case in dwarf galaxies, as discussed
in the next section.

\section{CO in dwarf galaxies and Blue Compact starbursts}

Much effort has been devoted to the detection of CO lines in
blue compact dwarf galaxies (BCDGs): their high star formation rates
is assumed to require a large H$_2$ content, and its detection will
help to understand the star formation mechanisms and efficiencies.
But the task has revealed very difficult, since the objects have low
masses, and low metallicity. Arnault et al. (1988) already suggested
that X was varying as Z$^{-1}$, and even as Z$^{-2}$ in a certain domain,
but this was contested (Sage et al 1992). Recent results improve
considerably the detection rate and the upper limits, because of the
technical progress in the receivers (Barone et al 1998,
Gondhalekar et al. 1998). They confirm the high dependence on
metallicity of the CO emission.

Taylor et al. (1998) have tried to detect in CO HI-rich dwarf galaxies,
in which the oxygen abundance was known. They have stringent upper
 limits on the most metal-poor galaxies, and conclude that the
conversion factor must be varying nonlinearly with metallicity, increasing 
sharply below $\sim$1/10 of the solar metallicity [12 + log (O/H) $\leq$ 7.9]
(cf figure 3).

\begin{figure}
\psfig{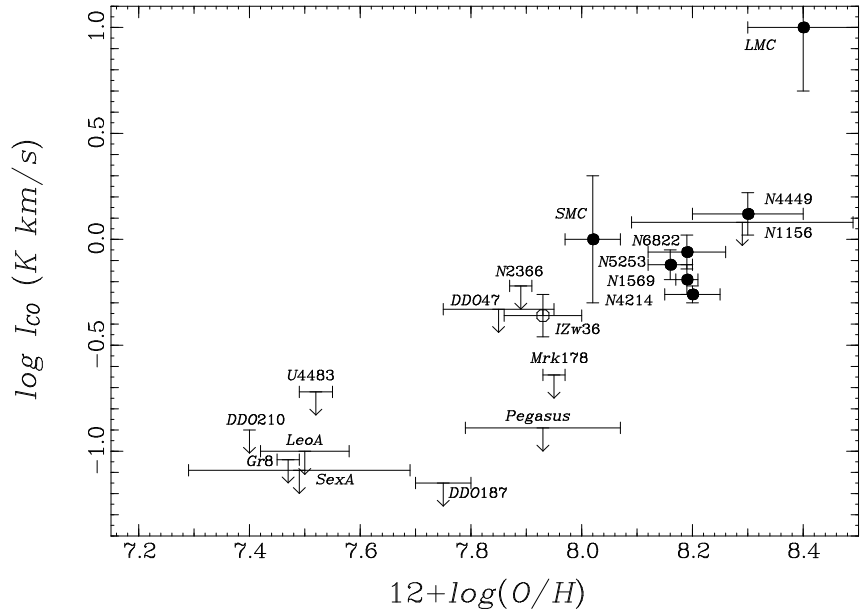}
\caption{ Dependence of the CO integrated intensity I$_{CO}$
(normally proportional to the average H$_2$ surface density N(H$_2$))
on the oxygen abundance, or metallicity (from Taylor et al. 1998).
}
\end{figure}

\section{Low surface brightness galaxies: LSB}

Low surface brightness galaxies may provide a clue to the galaxy
evolution processes: they appear to be unevolved galaxies, with large
gas fraction, which may have formed their
stars only on the second half of the Hubble time
(McGaugh \& de Blok 1997).
 Their metallicity is low, according to the correlation of
metallicity with surface brightness (Vila-Costas \& Edmunds 1992).
Some can have however, very large masses (and large rotational
velocities, implying large dark matter fractions).
 To tackle the mystery of their low-evolution rate,
it is of prime importance to know their molecular content, and their
total gas surface density. De Blok \& van der Hulst (1998) have
made a search for CO line emission in 3 late-type LSB galaxies:
they find a clear CO-emission deficiency, and conclude to a
molecular gas deficiency. They claim that the conversion factor X has
 not the same reasons to be higher than in normal late-type galaxies, since
the star formation rate is smaller than in HSB galaxies, implying a
lower UV flux, that does not photodissociate as much the CO
molecules. However, their derived upper limits  of the M(H$_2$)/M(HI) 
mass ratio fall in the same range as in HSB late-type galaxies; 
the question of their true molecular component is still open,
the more so as early-type LSB galaxies are detected in CO.
 Note that there exists a large scatter  of CO-emission
or H$_2$ content, even in normal galaxies, and this is not 
well understood: in M81 for example, a particularly low
H$_2$/HI interacting galaxy, with normal metallicity,
the molecular clouds appear to have different characteristics,
with a lower velocity dispersion at a given size (Brouillet 
et al. 1998).

Even with normal H$_2$/HI mass ratio (close to 1), the gas surface
density of those LSB galaxies is lower than critical
for gravitational instabilities, and that is sufficient to explain the low
efficiency of star formation (van der Hulst et al. 1993, van Zee et al.
1997). These low surface densities could come from
the poor environment and the lack of companions (Zaritsky \& Lorrimer 1993).

\section{Spiral Structure}

It becomes now possible to map in the CO lines galaxies at large-scale, and 
with high spatial resolution, to have an overview of the molecular
structure of a galaxy, with a high dynamical range.
With the On-The-Fly mapping procedure, Neininger et al. (1998a) have surveyed 
nearly half of the M31 disk in CO(1--0) with 23'' (90pc) resolution,
and they have detailed some remarkable regions, at 2'' resolution with
the IRAM interferometer. Apart from emphasizing the
fractal structure of the molecular gas over such large range of scales,
this study reveals a tight correlation between the CO arms and the
dust lanes, and also with the HI arms. At this scale, the kinematics
of the CO lines are more dominated by the star forming shells and the resulting
chaotic motions in the arms, than by large-scale streaming motions.
  In the global sense, the spiral structure of this galaxy is
not well determined, due to projection effects, and it is possible
that the ISM is concentrated more in a ring that in spiral arms
(see the ISOPHOT map from Haas et al 1998). 

The BIMA interferometer group has undertaken a survey of 44 nearby galaxies
(SONG), with large fields of view (mosaics of 7 fields of 3-4'),
with 7'' angular resolution (cf poster by Helfer et al., this conference).
There is a large range of CO morphologies detected, among spirals, 
bars and rings. The spiral structure of M51 is particularly detailed,
and the arm-interarm contrast deduced more exactly, including short-spacing.

\section{Centers of galaxies}

High-resolution CO maps of galaxy centers (at 100pc or less)
have been obtained with interferometers. One of the striking
features of the central H$_2$-component  structure is the strong
asymmetry observed, similar to what happens
in the center of the Milky Way: the gas distribution is
lopsided, and this appears to be the rule more than the exception.
  
One of the best example is the CO-rich M82 galaxy. Neininger et al
(1998b) have recently performed a 4'' resolution $^{13}$CO map with
the IRAM interferometer. The map shows the same gross features as the 
$^{12}$CO one made by Shen \& Lo (1995), i.e. a compact
central source, and two offset maxima, that could be the signature
of an edge-on ring. However, the central peak has a low $^{13}$CO/
$^{12}$CO  ratio, may be due to a large UV-photodissociation
(affecting selectively more the rarer isotopic species). 

There is a strong velocity gradient in the CO, which is also 
a general rule in spiral galaxies (cf Sofue et al. 1997).
The dynamical centre coincides with the IR peak and is shifted 6'' north-east 
of the compact $^{13}$CO source, emphasizing the lopsidedness.
The kinematics is also perturbed by star-formation shells, and around
the most luminous compact radio source in M 82, is identified a 
130 pc-wide bubble of molecular gas.

Another example is the edge-on warped galaxy NGC 5907, 
mapped in CO at 3'' resolution by Garcia-Burillo et al. (1997).
Spiral structure can be resolved in the center, which means that the galaxy is not completely
edge-on (or strongly warped even in the molecular component). Non-circular
motions are well explained by a bar rotating at a pattern speed of
$\Omega_b$ = 70 km/s/kpc (high enough to be that of a nuclear bar).
The velocity gradient is high in the center, resulting from a massive and
compact nuclear disk. This is interesting, since this galaxy could be classified
as an early-type according to its nuclear velocity gradient, but has no bulge
and is in fact classified as an Sc-Sd late-type. The molecular distribution
is significantly off-centered and lopsided also in this galaxy.

\begin{figure}
\psfig{figure=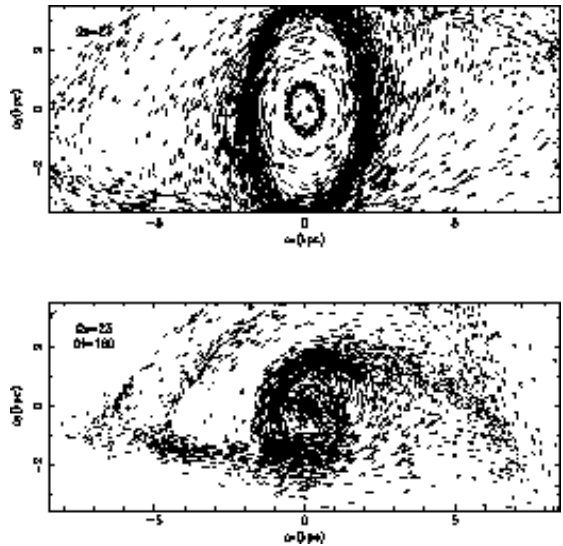,bbllx=15mm,bblly=19cm,bburx=85mm,bbury=25.5cm,width=8cm,angle=-0}
\caption{ Particle orbits of molecular clouds in a simulation with only one pattern speed
(top, $\Omega$ = 23 km/s/kpc), amd with two different pattern speeds for the 
two embedded bars (bottom, $\Omega$ = 23 and 160 km/s/kpc), in M100, 
from Garcia-Burillo et al. (1998a). Only in the second case, a nuclear spiral structure
is obtained, similar to the one observed in CO. } 
\end{figure}

\section{Double bars and nuclear disks}

Barred galaxies are conspicuous in general by their high concentrations of molecular gas
in the center. They often possess nuclear disks, or large flattened gas concentrations
in fast rotation, within the central 1kpc. This large molecular gas concentration
can trigger starbursts, sometimes confined in nuclear rings (hot spots in H$\alpha$).

The molecular distribution can have several morphologies, according to the
presence of zero, one or two inner Lindblad resonances in the center.
When there are Lindblad resonances, the gas accumulates in a spectacular
twin-peaks morphology (Kenney 1996), corresponding to the crossing
of the x1 orbits parallel to the bar, and the x2 orbits perpendicular to it,
inside the ILR, generally materialised by a nuclear ring.
 This typical structure is nicely  seen in NGC1530 mapped in CO by Reynaud \& Downes (1997):
the twin peaks are at the beginning of the characteristic thin dust lanes aligned with
and leading the bar. In rare cases, there is only a single peak, may be
due to the interaction with a companion (cf NGC5850, Leon et al 1999).
 When there is no ILR, in late-type galaxies for instance, or when 
the pattern speed is relatively high, there is no ring, and only a central
concentration (cf NGC 7479, Laine et al 1998).

Bars within bars is a frequently observed phenomenon, easy to see on color-color
plots, or in NIR images of galaxies (to avoid dust extinction). The presence of
an embedded bar has long been invoked as a mechanism to prolonge
the non-axisymmetry (and the resulting gravity torques) towards the center,
to drive the gas and fuel an AGN (Shlosman et al 1989). 
Simulations have described the numerical processus
leading to the formation of double bars (Friedli \& Martinet 1993, Combes 1994). 
In concentrating the mass towards the center, the first bar
modifies the inner rotation curve, and the precessing rate
($\Omega - \kappa/2$) of the $m=2$ elliptical orbits in the center increases to
large values. This widens the region between the two ILRs, and mass accumulate
on the perpendicular x2 orbits, which weakens the principal bar in this region.
This strong differential $\Omega - \kappa/2$
prevents the self-gravity from matching all precessing rates
in the center, and decoupling occurs: two bars rotating
at two different pattern speeds develop. 
Eventually, a too large mass accumulation into the center can
destroy the bar (Hasan et al 1993). To probe this scenario in double-bar
galaxies, it is important to know the pattern speeds of the two bars, and
derive the dynamical processus at play.

A prototype of double-bar galaxies is NGC4321, or M100 (cf Knapen \& Beckman 1996).
The study of its molecular cloud distribution has been done at high
resolution towards its central parts, including the nuclear bar
(Sakamoto et al. 1995, Garcia-Burillo et al 1998a). A small nuclear spiral structure
has been detected inside the nuclear ring made of the star-forming hot spots.
 This morphology requires  a model of nuclear bar with a very fast pattern speed
($\Omega$ = 160 km/s/kpc, see fig 4, from Garcia-Burillo et al. 1998a).

Sometimes, the gas in the center is also observed at high altitude above the plane
(cf NGC 891, 5907, or N 4013, Garcia-Burillo et al. 1999). These might be
accounted for by processes associated to the star-formation, more than purely
dynamical mechanisms. The only exception could be gas in retrograde orbits.

Garcia-Burillo et al. (1998b) report for the first time the detection of 
 a massive counterrotating molecular gas disk in the early-type spiral NGC3626.
The CO emission is concentrated in a compact nuclear disk of average radius r $\sim$ 12" (1.2 kpc),
rotating  in a sense opposite to that of the stars, and in the same sense as the HII and HI
gas (themselves counterrotating with respect to the stars).
There is no evidence of a violent starburst in the center of the galaxy, which
corresponds probably to the late stage of a merger.

\section{Molecular content of cluster galaxies}

Although many spiral galaxies in clusters are stripped from their HI gas,
 it has been established that they are not deficient in CO-emission, 
 and probably also H$_2$ (Kenney \& Young 1989 for Virgo; Boselli et al. 1995, 
Casoli et al. 1998 for Coma). This can be understood since the HI
is usually located in the outer parts of galaxies, where the gas is less
bound, and the tidal forces are larger; also the HI gas is more diffuse,
and easier to deplace by ram pressure. 

In Hickson compact groups, the molecular gas is not deficient
either, and even enhanced in tidally interacting galaxies
(Boselli et al. 1996, Leon et al. 1998, cf figure 5).
In a particular Hickson group, the Stephan's Quintet, Xu \& Tuffs (1998) have recently
found evidence of a starburst going on outside galaxies. From ISOCAM 15 $\mu$m
H$\alpha$ and NIR observations, they identified an outstanding bright source
about 25kpc away from the neighbouring galaxies, containing very young stars,
and corresponding to an SFR of about 0.7 M$_\odot$/yr. They propose that the 
starburst is triggered by the collision between a fast galaxy an the IGM; alternatively
this could be the formation of a tidal dwarf, through  gravitational instabilities
in a tidal tail (cf Duc \& Mirabel 1998). Molecular gas (through CO emission) is difficult to detect
so far from the nuclei (possibly because of metallicity gradients), and the
detection in tidal tails are rare, like that in the NGC 2782 (Arp 215) tail
by Smith et al. (1998).

\begin{figure}
\psfig{figure=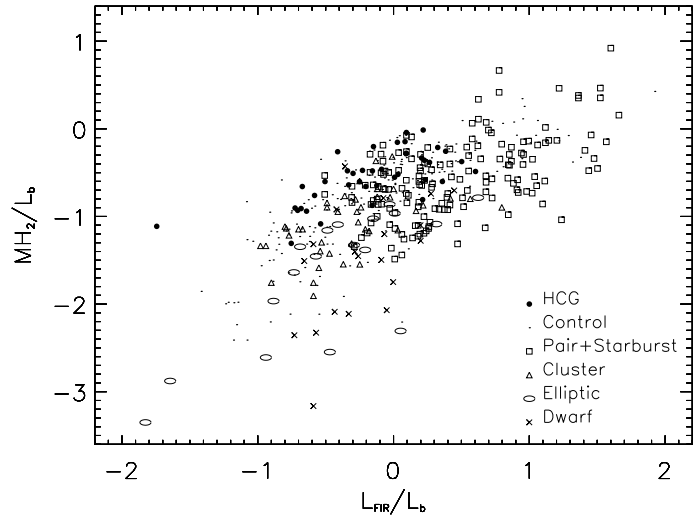,bbllx=0cm,bblly=16cm,bburx=10cm,bbury=28cm,width=10cm,angle=90}
\caption{ M(H$_2$) versus L(FIR) normalised to Blue luminosity L$_b$ for galaxies in compact
groups HCG, compared to other samples (Control, Pairs and starbursts, Cluster, Elliptical
and Dwarfs) from Leon et al. (1998). }
\end{figure}

\section{Ultra-luminous IRAS galaxies}

A new survey of ultra-luminous infrared galaxies with millimeter interferometer
reveals that the molecular gas is confined in compact rotating nuclear disks or
rings (Downes \& Solomon 1998). The constraint that the gas mas is smaller
than the dynamical mass imposes an excitation model in which the CO lines
are only moderately optically thick ($\tau = 4-10$) and subthermally
excited, so that the CO-to-H$_2$ conversion ratio is about 5 times less
than standard. In that case, the fraction gas-to-dynamical mass is 15\%.
The surface density of gas, however, is in average $\mu_g$/$\mu_{tot}$ = 30\%.
 In some cases, the CO position-velocity diagrams clearly shows a ring,
with a central gap. In the particular case of Arp 220, there appears to be
two bright sources embedded in a central nuclear disk, which are 
compact extreme starburst regions, more likely than the premerger nuclei, as was
previously thought. 

An often debated question is whether the huge far infrared emission of
the ultra-luminous galaxies are due to a starburst or a monster (AGN).
It is frequent that both are present simultaneously, since they are both
the results of huge mass accumulation in the centers of galaxies.
Using ISO data, and diagnostic diagrams involving the ratio of high-to-low
excitation mid-IR emission lines, together with the strength of the 7.7 $\mu$m
``PAH'' feature, Genzel et al. (1998) conclude that 
the far infrared emission appears to be powered predominantly by starbursts:
70\%-80\%  are predominantly powered by recently formed massive stars, and 
20\%-30\% are powered by a central AGN. Very high extinctions are measured 
towards these star-forming regions, supporting the high H$_2$ column
densities derived from CO observations. In these objects, the active region
is always sub-kpc in size.

\section{Galaxies at high redshift}

One of the most exciting results of these last years is the detection of galaxies
at larger and larger redshifts, allowing to tackle the evolution and history of
star formation in the Universe. After the first discovery in CO lines of
an object at $ z > 2$, the ultraluminous galaxy IRAS 10214+4724
(Brown \& van den Bout 1992, Solomon et al 1992), there has been an
extended search for more early starbursts, that resulted in the
discovery of about 10 objects at high $z$ in CO lines (cf Table \ref{COdata}).

\begin{table}[h]
\caption[ ]{CO data for high redshift objects}
\begin{flushleft}
\begin{tabular}{lllclcl}  \hline
Source    & $z$   &  CO  & S  & $\Delta$V& MH$_2$   & Ref  \\
          &       &line  & mJy  & km/s & 10$^{10}$ M$_\odot$    &           \\
\hline
F10214+4724 & 2.285 & 3-2  & 18 & 230  & 2$^*$      &  1   \\
53W002      & 2.394 & 3-2  &  3 & 540  & 7          &  2   \\
H 1413+117  & 2.558 & 3-2  & 23 & 330  & 2-6        &  3   \\
SMM 14011+0252&2.565& 3-2  & 13 & 200  & 5$^*$      &  4   \\
MG 0414+0534& 2.639 & 3-2  &  4 & 580  & 5$^*$      &  5   \\
SMM 02399-0136&2.808& 3-2  &  4 & 710  & 8$^*$      &  6   \\
APM 08279+5255&3.911& 4-3  &  6 & 400  & 0.3$^*$    &  7   \\
BR 1335-0414& 4.407 & 5-4  &  7 & 420  & 10         &  8   \\
BR 1202-0725& 4.690 & 5-4  &  8 & 320  & 10         &  9   \\
\hline 
\end{tabular}
\end{flushleft}
$^*$ corrected for magnification, when estimated\\
Masses have been rescaled to $H_0$ = 75km/s/Mpc. When multiple images
are resolved, the flux corresponds to their sum\\
(1) Solomon et al. (1992a), Downes et al. (1995); (2) Scoville et al. (1997b); (3
) Barvainis et al. (1994, 1997); (4) Frayer et al. (1999);
(5) Barvainis et al. (1998); (6) Frayer et al. (1998); (7)
Downes et al. (1998); (8) Guilloteau et al. (1997); (9) Omont et al. (1996a)
\label{COdata}
\end{table}

In fact the majority of these objects (if not all) are amplified by gravitational
lenses, and this explains why they are detectable at all (see 
figure 6).  The amplification is very helpful to detect these remote
objects, but the drawbacks are significant  uncertainties in the amplification
factors, and therefore on the total molecular content. The excitation
of the gas is also uncertain, since the various CO lines emission may have 
different spatial extents and different resulting amplifications.

One strategy to search for CO lines in high-$z$ galaxies has been to select
objects already detected in the far-infrared or submm dust emission.
Indeed, all objects in Table \ref{COdata} have been first detected 
in continuum, some (the SMM) have been discovered in blank field searches
with the SCUBA bolometer on JCMT (Hawaii) by Smail et al. (1997).
Since the emission of the dust is varying as $\nu^4$ with the frequency $\nu$
in the millimeter range (until the maximum near 60$\mu$m), it becomes
easier to detect galaxies at $z$ = 5 than $z$ = 1 (cf Blain \& Longair 1993).
 The millimeter domain is then a privileged one to follow the star-formation
history as a function of redshift, and several surveys have been undertaken.
Searches toward the Hubble Deep Field-North
(Hughes et al 1998), and towards the Lockman hole and SSA13 (Barger et al 1998),
have found a few sources, revealing an increase of starbursting galaxies with redshift.
They correspond  by extrapolation to a density of
800 per square degree, above 3 mJy at 850 $\mu$m. This already can account for
50\% of the cosmic infra-red background (CIRB), that has been estimated by
Puget et al (1996) and Hauser et al (1998) from COBE data.
Similar conclusions have been reached towards the CFRS fields by Eales et al. (1999).

These first results show the potentiality of the millimeter domain, already with 
the present instruments. With the fore-coming next generation of mm telescopes,
which will yield a factor 10-20 increase in sensitivity, it will be possible to
detect not only huge starbursts but more ordinary galaxies at high redshift
(cf Combes et al 1999).

\begin{figure}
\psfig{figure=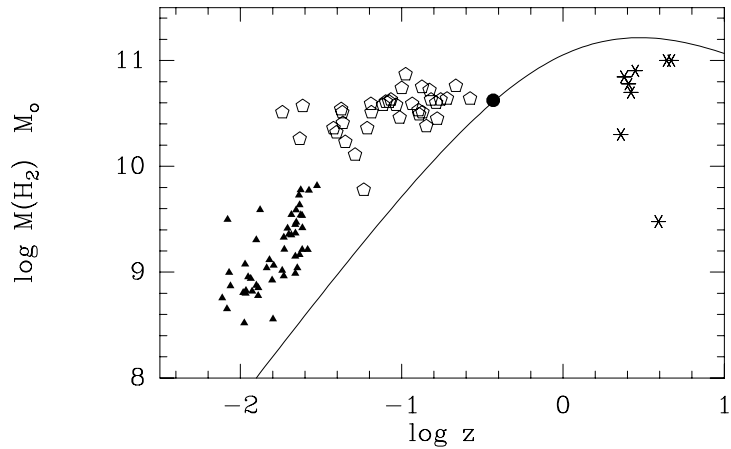,bbllx=2cm,bblly=1cm,bburx=12cm,bbury=16cm,width=10cm,angle=-90}
\caption{H$_2$ masses for the CO-detected objects at high redshift (filled
stars), compared to the ultra-luminous-IR sample of Solomon et al (1997,
open pentagons), and to the Coma supercluster sample from Casoli et
al (1996, filled triangles).
There is no detected object between 0.3 and 2.2 in redshift, except the quasar
3c48, marked as a filled dot (Scoville et al 1993, Wink et al 1997). The curve
indicates the 3$\sigma$ detection limit of I(CO) = 1 K km/s at the IRAM-30m
telescope (equivalent to an rms of 1mK, with an assumed $\Delta V$ = 300km/s).
The points at high $z$ can be detected well below this limit, since they are
gravitationally amplified. }
\end{figure}

\end{document}